\documentclass[aps,prd,onecolumn,groupedaddress,showpacs,nofootinbib,amssymb]{revtex4}
\usepackage[dvips]{graphicx}
\usepackage{amssymb}
\usepackage{amsmath}
\usepackage{graphicx,,color}
\usepackage{amsfonts}
\usepackage{bm}
\usepackage{cancel}
\usepackage{comment}
\usepackage[dvipsnames]{xcolor}
\usepackage{makecell}

\allowdisplaybreaks[4]

\begin{document}

\tolerance=5000

\title{Strong Gravity Effects on $\mathcal{R}^2$-corrected Single Scalar Field Inflation and Compatibility with the ACT Data}
\author{V.K. Oikonomou$^{1,2}$}\email{voikonomou@gapps.auth.gr;v.k.oikonomou1979@gmail.com}
\affiliation{$^{1)}$Department of Physics, Aristotle University of
Thessaloniki, Thessaloniki 54124, Greece} \affiliation{$^{2)}$L.N.
Gumilyov Eurasian National University - Astana, 010008,
Kazakhstan}

\begin{abstract}
In this work we introduce the rescaled $\mathcal{R}^2$-corrected
minimally coupled scalar field theory, a theory that contains
minimal quantum corrections of the single scalar field Lagrangian.
We develop the theoretical framework in the Jordan frame where the
baryons geodesics are free fall geodesics and we do not treat the
theory as a two scalar field theory in the Einstein frame. The
theoretical framework can be reduced to a single scalar field
theory framework by using a perturbative expansion at the level of
the field equations, making the resulting theory easy to tackle
analytically. The first two quantum corrections contain two terms,
a linear $\sim \mathcal{R}$ and a quadratic term $\sim
\mathcal{R}^2$. The effect of the linear term alters the
Einstein-Hilbert term, making the resulting theory a rescaled
version of Einstein-Hilbert gravity. Due to the presence of the
rescaled Einstein-Hilbert term $\sim \lambda
\frac{\mathcal{R}}{16\pi G}$, the gravitational constant will no
longer be that of Newton's, but a rescaled one $\frac{G}{\lambda}$
and hence gravity can be stronger primordially, or even weaker.
The perspective of having stronger gravity primordially, is
compatible with intuition, since one expects a stronger gravity
primordially, but having a weaker gravity for some reason is not
prohibited theoretically. The attribute of our theoretical
framework is that it allows a stronger gravity primordially, which
returns to ordinary gravity as the curvature of the Universe
decreases. We examine the effects of the quantum terms on several
mainstream scalar field inflationary potentials, such as hybrid
inflation, monomial inflation and power-law inflation, and we
demonstrate that the hybrid inflation and power-law inflation,
which are 3-parameter models, are compatible with the Atacama
Cosmology Telescope data in the strong gravity limit, while the
monomial inflation, which is a 4-parameter model, can be
compatible with the Atacama Cosmology Telescope data in the weak
gravity limit, however the model suffers from perturbative
expansion break down during inflation.
\end{abstract}

\maketitle

\section{Introduction}

The primordial era of our Universe is undoubtedly the most
mysterious and at the same time the most interesting era from a
theoretical point of view. The reason is that any Ultraviolet (UV)
completion of the Standard Model affects exactly this era. In the
absence of any solid result from the Large Hadron Collider
regarding the UV behavior of the Standard Model, the scientific
community has turned its focus on the observations coming from the
sky and specifically the Cosmic Microwave Background (CMB). The
CMB may encode the undiscovered, to the moment UV, completions of
our Universe, in the relics of a theory that is called inflation.
Inflation \cite{inflation1,inflation2,inflation3,inflation4} is a
hypothetical post-Planck era of our Universe, which may have
imprints of the Planck era in terms of lower order effective terms
that affect the evolution of our Universe. Inflation itself is a
classical theory and the Universe is itself is four dimensional
and classical, meaning that no severe quantum effects occur during
this era. However, as we already mentioned inflation may have some
quantum effects in the form of one loop correction terms  which
may be present in the inflationary Lagrangian. The skopos of
theoretical physicist is to find evidence for these quantum
effects on the CMB. Inflation has been put to the test for at
least three decades so far, and will further be put to the test in
the near future, by CMB-based experiments like the Simons
Observatory \cite{SimonsObservatory:2019qwx} and the LiteBird
\cite{LiteBIRD:2022cnt}, but also from future gravitational wave
experiments
\cite{Hild:2010id,Baker:2019nia,Smith:2019wny,Crowder:2005nr,Smith:2016jqs,Seto:2001qf,Kawamura:2020pcg,Bull:2018lat,LISACosmologyWorkingGroup:2022jok}.

Now the debate among theorists is for the terms that should be the
encoding of the UV era of our Universe on the classical
inflationary era. A plain and simple way of thinking, aligned with
Occam's razor approach on problems, the UV completion should
contain higher orders of the Ricci scalar $\mathcal{R}$. The
reason for this is simple, the Einstein-Hilbert Lagrangian
contains a linear term of the Ricci scalar $\mathcal{R}$, thus any
UV completion should probably contain higher order terms. Thus
$f(\mathcal{R})$ gravity enjoys undoubtedly an elevated role among
many other modified gravities
\cite{reviews1,reviews2,reviews3,reviews4}. The question is then,
which $f(\mathcal{R})$ terms should be considered to contribute to
the UV completion of classical theories, like general relativity?
One again must think in a simple context. What we have found so
far is the Higgs, a spin zero scalar particle and also we have
general relativity. Thus by assuming that a scalar field realizes
the inflationary era, which is standard in the inflationary
literature, the higher order curvature terms should affect the
scalar field inflationary Lagrangian. Indeed, as we show later,
the first corrections to the scalar field Lagrangian, when the
scalar field is considered in its vacuum configuration, contain
indeed $f(\mathcal{R})$ terms, the first of which is a linear term
in the Ricci scalar and the second is an $\mathcal{R}^2$ term
\cite{Codello:2015mba}. Such $\mathcal{R}^2$ corrected scalar
field Lagrangians are met in the literature quite frequently
\cite{Ema:2017rqn,Ema:2020evi,Ivanov:2021ily,Gottlober:1993hp,delaCruz-Dombriz:2016bjj,Enckell:2018uic,Kubo:2020fdd,Gorbunov:2018llf,Calmet:2016fsr,Oikonomou:2021msx,Oikonomou:2022bqb,Pi:2017gih,Salvio:2016vxi,Kannike:2015apa,Salvio:2015kka},
however the general approach in these studies is to treat the
theory as having two scalar fields and in the Einstein frame, by
conformally transforming the $\mathcal{R}^2$ term.

In this work we aim to consider the $\mathcal{R}^2$ corrected
minimally coupled scalar field theory in the Jordan frame
directly. The attribute of our approach is that baryons in the
Jordan frame follow free fall geodesics and the particles are not
coupled to the metric in any way. Geodesics of particles are
free-fall geodesics. Using the first two quantum corrections of a
minimally coupled scalar field Lagrangian, that is a linear $\sim
\mathcal{R}$ and a quadratic term $\sim \mathcal{R}^2$, we shall
formulate the new theory in such a way so that the resulting
framework is a perturbative single scalar field framework in the
Jordan frame. Also we shall seek the perturbative effects of the
quantum corrections $\sim \mathcal{R}$ and of the quadratic term
$\sim \mathcal{R}^2$ on the single scalar field Lagrangian. One of
the interesting features of the resulting theory is a rescaling of
the standard Einstein-Hilbert term of general relativity. Such a
rescale is effective only primordially, when the curvature of the
Universe is high and more importantly, it affects Newton's
gravitational constant directly. The reasoning is simple, the
rescaled Einstein-Hilbert term $\sim \lambda
\frac{\mathcal{R}}{16\pi G}$, will have a rescaled gravitational
constant $\frac{G}{\lambda}$ and thus will become weaker or
stronger than Einstein's gravity. Theoretically speaking, there is
no reason for the gravitational constant $G$ to be the same
primordially and at present day. In fact, effective quantum
effects may affect the gravitational constant directly through
effective terms, thus gravity itself may be stronger or weaker
during inflation. Intuitively, gravity must be stronger
primordially, but this is just intuition imposed by the fact that
primordially the Universe was small and gravitational effects
should be in principle stronger. Recently, the Atacama Cosmology
Telescope (ACT) data \cite{ACT:2025fju,ACT:2025tim} combined with
the DESI data \cite{DESI:2024uvr}, made the scientific community
to reconsider the benchmark primordial theory of our Universe,
that is inflation, since the ACT data indicated that the scalar
spectral index of the primordial curvature perturbations, is in at
least 2$\sigma$ discordance with the Planck data
\cite{Planck:2018jri}. Specifically, the scalar spectral index of
the scalar perturbations is found by the ACT data to be
constrained as follows,
\begin{equation}\label{act}
n_{\mathcal{S}}=0.9743 \pm
0.0034,\,\,\,\frac{\mathrm{d}n_{\mathcal{S}}}{\mathrm{d}\ln
k}=0.0062 \pm 0.0052\, .
\end{equation}
Moreover, the updated Planck constraint on the tensor-to-scalar
ratio is \cite{BICEP:2021xfz},
\begin{equation}\label{planck}
r<0.036\, ,
\end{equation}
at $95\%$ confidence. Thus, inflation is significantly different
under the prism of the ACT data, and this result already created a
large stream of articles aiming to align with the ACT data
\cite{Kallosh:2025rni,Gao:2025onc,Liu:2025qca,Yogesh:2025wak,Yi:2025dms,Peng:2025bws,Yin:2025rrs,Byrnes:2025kit,Wolf:2025ecy,Aoki:2025wld,Gao:2025viy,Zahoor:2025nuq,Ferreira:2025lrd,Mohammadi:2025gbu,Choudhury:2025vso,Odintsov:2025wai,Q:2025ycf,Zhu:2025twm,Kouniatalis:2025orn,Hai:2025wvs,Dioguardi:2025vci,Yuennan:2025kde,Kuralkar:2025zxr,Kuralkar:2025hoz,Aoki:2025ywt},
which however must be used with caution \cite{Ferreira:2025lrd}.
In this work we shall show that the rescaled
$\mathcal{R}^2$-corrected minimally coupled scalar field theory
may be easily compatible with the ACT data. We shall develop the
framework of rescaled $\mathcal{R}^2$-corrected minimally coupled
scalar field theory in the Jordan frame and we shall formulate the
theory as a perturbative single scalar field theory. We shall also
test the theory using three mainstream scalar field models, two
3-parameter models and one 4-parameter model. We shall demonstrate
that all the strong gravity models produce a viable and
self-consistent inflationary era compatible with the ACT data. The
weak gravity 4-parameter model produces a viable inflationary era,
compatible with the ACT data, but the model has inconsistencies,
breaking the perturbation expansion inherently.

This work is organized as follows: In section II we present the
general theoretical framework of rescaled
$\mathcal{R}^2$-corrected minimally coupled scalar field theory.
We discuss how the quantum corrections can be used to perform a
perturbation expansion at the equations of motion level and we
discuss how the theory can be formulated in terms of a single
scalar field theory. We derive the slow-roll indices and analytic
expressions for the observational indices of inflation. In section
III we analyze two strong gravity and one weak gravity mainstream
single scalar field potentials and indicate that all the models
are compatible with the ACT data, however, the weak gravity has
inherent inconsistencies, due to the fact that the perturbative
expansion of the quantum corrections breaks during the
inflationary era. The conclusions follow in the end of the
article.

\section{Formalism of Rescaled Canonical Scalar Field Inflation with $R^2$ Corrections}

Let us discuss why one expects higher order curvature corrections
in a scalar field theory in general. We consider the most general
scalar field action, which contains at most two derivatives,
\begin{equation}\label{generalscalarfieldaction}
\mathcal{S}_{\phi}=\int
\mathrm{d}^4x\sqrt{-g}\left(\frac{1}{2}Z(\phi)g^{\mu
\nu}\partial_{\mu}\phi
\partial_{\nu}\phi-\mathcal{V}(\phi)+h(\phi)\mathcal{R}
\right)\, .
\end{equation}
When the scalar field is considered in its vacuum configuration,
the coupling of the scalar field to gravity must either be
conformal or minimal. In this work we shall consider minimally
coupled scalar field theory, so we take $Z(\phi)=-1$ and
$h(\phi)=1$ in Eq. (\ref{generalscalarfieldaction}). The quantum
corrected action of the single scalar field action
(\ref{generalscalarfieldaction}) contains the following terms
\cite{Codello:2015mba},
\begin{align}\label{quantumaction}
&\mathcal{S}_{eff}=\int
\mathrm{d}^4x\sqrt{-g}\Big{(}\Lambda_1+\Lambda_2
\mathcal{R}+\Lambda_3\mathcal{R}^2+\Lambda_4 \mathcal{R}_{\mu
\nu}\mathcal{R}^{\mu \nu}+\Lambda_5 \mathcal{R}_{\mu \nu \alpha
\theta}\mathcal{R}^{\mu \nu \alpha \theta}+\Lambda_6 \square
\mathcal{R}\\ \notag &
+\Lambda_7\mathcal{R}\square\mathcal{R}+\Lambda_8 \mathcal{R}_{\mu
\nu}\square \mathcal{R}^{\mu
\nu}+\Lambda_9\mathcal{R}^3+\mathcal{O}(\partial^8)+...\Big{)}\, ,
\end{align}
with the parameters $\Lambda_i$, $i=1,2,...,6$ being appropriate
dimensionful constants. In this work we shall consider leading
$\mathcal{R}$ and $\mathcal{R}^2$ corrections and we shall
investigate the effect of these two terms on single scalar field
inflationary dynamics. The gravitational action we shall consider
is the following,
\begin{equation}
\label{action} \centering
\mathcal{S}=\int{d^4x\sqrt{-g}\left(\frac{ \lambda
\mathcal{R}+\frac{\mathcal{R}^2}{36M^2}}{2\kappa^2}-\frac{1}{2}g^{\mu
\nu}\partial_{\mu}\phi
\partial_{\nu}\phi-\mathcal{V}(\phi)\right)}\, ,
\end{equation}
where $\kappa^2=8\pi G=\frac{1}{M_p^2}$ and $M_p$ stands for the
reduced Planck mass, where $M$ is a mass scale to be determined
later on when specific scalar field inflation models will be
considered. Let us note here that the value of the mass scale $M$
is not the one of the standard $\mathcal{R}^2$ model
\cite{Starobinsky:1980te,Bezrukov:2007ep}. As it can be seen, the
action (\ref{action}) is a gravitational theory with a rescaled
Einstein-Hilbert term $\mathcal{R}$. The reason and motivation for
this rescaled theory is sourced in the quantum effects that we
took into account for the scalar field theory. Specifically, it is
not certain that primordially the strength of the gravitational
field is the same with its present day strength. What does
rescaled gravity actually do is that it rescales Newton's
gravitational constant. The resulting theory can be a stronger or
weaker gravity compared to the Einstein-Hilbert gravity, purely
motivated by quantum gravitational effects on the single scalar
field Lagrangian. There is another reason that a rescaled
gravitational theory might emerge. Specifically, if
$f(\mathcal{R})$ gravity controls the gravitational dynamics of
the Universe as a whole, then the complete $f(\mathcal{R})$
gravity might contain terms that primordially can effectively
alter Newton's gravitational constant. These $f(\mathcal{R})$
theories can usually provide a unified description of inflation
and the dark energy era, and have been formally studied in Ref.
\cite{Oikonomou:2025qub}. Most of them contain exponential terms
which primordially can yield a rescaled Einstein-Hilbert gravity
\cite{Oikonomou:2025qub}. An example of this sort, is the
following,
\begin{equation}\label{frini}
f(\mathcal{R})=\mathcal{R}-\gamma  \theta  \Lambda -\theta
\mathcal{R} \exp \left(-\frac{\gamma \Lambda
}{\mathcal{R}}\right)-\frac{\Lambda
\left(\frac{\mathcal{R}}{m_s^2}\right)^{\delta }}{\zeta }\, ,
\end{equation}
where $\Lambda$ is the cosmological constant, and
$m_s^2=\frac{\kappa^2 \rho_m^{(0)}}{3}$, where $\rho_m^{(0)}$
denotes the energy density of cold dark matter at present day. The
above, in the large curvature limit yields the following form,
\begin{equation}\label{expapprox}
\theta  \mathcal{R} \exp \left(-\frac{\gamma  \Lambda
}{\mathcal{R}}\right)\simeq -\gamma \theta  \Lambda -\frac{\gamma
^3 \theta \Lambda^3}{6 \mathcal{R}^2}+\frac{\gamma ^2 \theta
\Lambda ^2}{2 \mathcal{R}}+\theta \mathcal{R}\, ,
\end{equation}
therefore, the complete effective action when the curvature is
high takes the form,
\begin{equation}\label{effectiveaction}
\mathcal{S}=\int
d^4x\sqrt{-g}\left(\frac{1}{2\kappa^2}\left(\lambda \mathcal{R}+
\frac{\gamma ^3 \theta \Lambda ^3}{6 \mathcal{R}^2}-\frac{\gamma
^2 \theta \Lambda ^2}{2 \mathcal{R}}-\frac{\Lambda}{\zeta
}\left(\frac{\mathcal{R}}{m_s^2}\right)^{\delta
}+\mathcal{O}(1/\mathcal{R}^3)+...\right)-\frac{1}{2}g^{\mu\nu}\nabla_\mu\phi\nabla_\nu\phi-\mathcal{V}(\phi)\right)\,
,
\end{equation}
where $\lambda=1-\theta$. For all the rescaled gravity scenarios,
there is the peril of affecting Big Bang Nucleosynthesis physics,
due to altering Newton's gravitational constant. This is not
however a true peril for the rescaled theories, because such
effective theories emerge in the high curvature regime, during
inflation, and thus do not affect the low curvature regime of
later phases of our Universe's evolution.

Let us consider a flat Friedmann-Robertson-Walker (FRW) spacetime,
\begin{equation}
    \centering\label{metric}
    \mathrm{d}s^2 = - \mathrm{d} t^2 + a(t) \sum_{i = 1}^3 \mathrm{d} x_i^2,
\end{equation}
and let us extract the field equations for the rescaled
$\mathcal{R}^2$ gravitational action of Eq. (\ref{action}), which
are,
\begin{equation} \label{Friedmann}
    3 f_{\mathcal{R}} \mathcal{H}^2=\frac{R f_{\mathcal{R}} -f}{2}-3\mathcal{H} \dot F_{\mathcal{R}}+\kappa^2\big(\frac{1}{2}\dot\phi^2+\mathcal{V}(\phi)\big) \ ,
\end{equation}
\begin{equation} \label{Raychad}
    -2 f_{\mathcal{R}} \dot{\mathcal{H}} = \kappa^2 \dot\phi^2 + \ddot f_{\mathcal{R}} -\mathcal{H} \dot f_{\mathcal{R}} \ ,
\end{equation}
\begin{equation} \label{fieldeqmotion}
   \ddot\phi+3\mathcal{H}\dot\phi+\mathcal{V}'=0 \ ,
\end{equation}
with the ``dot'' denoting derivatives with respect to the cosmic
time, the ``prime'' denotes derivative with respect to the scalar
field $\phi$, and also $f_{\mathcal{R}}=\frac{\partial f}{\partial
\mathcal{R}}$. For $f(\mathcal{R})=\lambda
\mathcal{R}+\frac{\mathcal{R}^2}{36M^2}$ and in addition that for
a flat FRW metric, the Ricci scalar and its derivative are,
\begin{equation}\label{ricciscalar}
    \mathcal{R} =12 \mathcal{H}^2 + 6\dot{\mathcal{H}} \, , \ \dot{\mathcal{R}}=24\mathcal{H}\dot{\mathcal{H}} +6\ddot{\mathcal{H}},
\end{equation}
the Friedmann and Raychaudhuri equations take the following form,
\begin{equation} \label{Friedman3}
    3\lambda\mathcal{H}^2+\frac{\mathcal{H}^2}{M^2}\dot{\mathcal{H}}=\kappa^2\mathcal{V}(\phi) ,
\end{equation}
\begin{equation} \label{Raycha3}
    -2 \lambda \dot{\mathcal{H}}-\frac{2}{M^2}\dot{\mathcal{H}}^2 =\kappa^2 \dot \phi^2,
\end{equation}
\begin{equation} \label{dotphi}
   \dot \phi \simeq - \frac{\mathcal{V}'}{3\mathcal{H}} .
\end{equation}
where we assumed that the slow-roll conditions apply for the
inflationary era,
\begin{equation}\label{slowrollH}
    \dot{\mathcal{H}} \ll \mathcal{H}^2 \ , \ \ddot{\mathcal{H}} \ll \mathcal{H} \dot{\mathcal{H}},
\end{equation}
and that the following slow-roll motivated conditions also  hold
true,
\begin{equation}\label{approxH}
    \frac{\dot{\mathcal{H}}^2}{M^2} \ll \mathcal{H}^2,\,\,\,\frac{\dot{\mathcal{H}}^2}{M^2} \ll
    \mathcal{V}(\phi)\, .
\end{equation}
The Raychaudhuri equation is  a second order polynomial equation
with respect to the derivative of the Hubble rate
$\dot{\mathcal{H}}$, so upon solving it we obtain,
\begin{equation}\label{dothanalyticsol}
\dot{\mathcal{H}}=\frac{-M^2\lambda+M\sqrt{M^2\lambda^2-2\dot{\phi}^2\kappa^2}}{2}\,
.
\end{equation}
We shall make the following assumption,
\begin{equation}\label{taylorphi}
    \frac{2\kappa^2 \dot \phi^2}{M^2} \ll 1\, ,
\end{equation}
which will be tested for all the models we shall consider in this
work. In view of Eq. (\ref{taylorphi}), the Friedmann and
Raychaudhuri equation at leading order become,
\begin{equation} \label{Friedman}
    \lambda \mathcal{H}^2 \simeq \frac{\kappa^2 \mathcal{V}(\phi)}{3}+\mathcal{O}(\frac{\kappa^2 \dot{\phi}^2}{2M^2}\mathcal{H}^2),
\end{equation}
\begin{equation} \label{Raycha}
     \dot{\mathcal{H}} \simeq -\frac{\kappa^2 \dot{\phi}^2}{2\lambda}  -\frac{\kappa^4
     \dot{\phi}^4}{4\lambda^3 M^2}\, ,
\end{equation}
where we used the expansion,
\begin{equation}\label{expansionref}
\sqrt{\lambda ^2-x}\simeq \lambda -\frac{x^2}{8 \lambda
^3}-\frac{x}{2 \lambda }\, .
\end{equation}
The above field equations Eqs. (\ref{Friedman}), (\ref{Raycha})
and (\ref{dotphi}) will be the core of our analysis. These
equations contain the leading order quantum effects primordially
and hereafter we shall examine the effects of the quantum
corrections on the inflationary dynamics. The slow-roll indices
for a $f(\mathcal{R},\phi)$ theory are \cite{Hwang:2005hb},
\begin{align}\label{slowrollindices}
\epsilon_1=-\frac{\dot{\mathcal{H}}}{\mathcal{H}^2},\,\,\,\epsilon_2=\frac{\ddot
\phi }{\mathcal{H} \dot \phi},\,\,\,
\epsilon_3=\frac{\dot f_{\mathcal{R}} }{2\mathcal{H}f_{\mathcal{R}}},\,\,\,\epsilon_4=\frac{\dot E}{2\mathcal{H}E}\, ,\\
\end{align}
with,
\begin{equation}\label{E1}
    E=f_{\mathcal{R}}+\frac{3 \dot f_{\mathcal{R}}^2}{3\kappa^2\dot
    \phi^2}\, .
\end{equation}
In view of Eqs. (\ref{Friedman}) and (\ref{Raycha}), the first
slow-roll index $\epsilon_1$ takes the following form,
\begin{equation}\label{e1}
    \epsilon_1=\frac{\lambda}{2\kappa^2}\left(\frac{\mathcal{V}'}{\mathcal{V}}\right)^2 +\frac{1}{12M^2}{\left(\frac{\mathcal{V}'}{\mathcal{V}}\right)}^2\frac{\mathcal{V}'^2}{\mathcal{V}}\, .
\end{equation}
Also, after some algebra, we obtain the second slow-roll index,
\begin{equation}\label{e2}
    \epsilon_2=-\frac{\mathcal{V}''}{\kappa^2 \mathcal{V}}+\epsilon_1 \, ,
\end{equation}
and in addition, the slow-roll index $\epsilon_3$ is equal to,
\begin{equation}\label{e3}
    \epsilon_3=\frac{\epsilon_1}{-1-\frac{3\lambda M^2}{2\mathcal{H}^2}+\frac{\epsilon_1}{2}} \, .
\end{equation}
Furthermore, by calculating $E$ and $\dot E$,
\begin{equation}\label{E}
    E=\frac{\lambda}{\kappa^2}+\frac{4\mathcal{H}^2}{3M^2\kappa^2}+\frac{3}{2\kappa^4\dot{\phi}^2}\left(\frac{4\mathcal{H}\dot{\mathcal{H}}}{3M^2} \right)^2\, ,
\end{equation}
\begin{equation}\label{dotE}
    \dot E=\frac{4 \mathcal{H} \dot{\mathcal{H}}}{3M^2} + \frac{16}{\kappa^4\dot{\phi}^4M^4}\left(\mathcal{H}\dot{\mathcal{H}}^3\dot{\phi}^2-\mathcal{H}^2\dot{\mathcal{H}}^2\dot{\phi}\ddot{\phi} \right),
\end{equation}
we can obtain the slow-roll index $\epsilon_4$. Now, given the
slow-roll indices, one may obtain the observational indices of
inflation, expressed in terms of the slow-roll indices. The
spectral index of the scalar curvature perturbations is equal to
\cite{Hwang:2005hb},
\begin{equation}\label{ns}
    n_\mathcal{S}= 1 - \frac{4\epsilon_1 + 2\epsilon_2 -2 \epsilon_3 + 2 \epsilon_4}{
 1 -\epsilon_1} .
\end{equation}
Regarding the tensor-to-scalar ratio, for the
$f(\mathcal{R},\phi)$ theory at hand is equal to,
\begin{equation}\label{r}
    r=16(\epsilon_1 + \epsilon_3)\, .
\end{equation}
Also an important quantity which is necessary for the analytical
treatment of the inflationary dynamics is the $e$-foldings number,
which for the $f(\mathcal{R},\phi)$ theory at hand acquires the
following form,
\begin{equation}\label{N1}
    N = \int_{t_i} ^{t_f} \mathcal{H} d t=\int_{\phi_i} ^{\phi_f} \frac{\mathcal{H}}{\dot \phi} d \phi ,
\end{equation}
where we used Eqs. (\ref{ns})-(\ref{r}). Note that $t_i$ and
$\phi_i$ denote the time instance and the scalar field value at
the beginning of the inflationary era, when the first horizon
crossing of the modes occurred, and $t_f$ and $\phi_f$ denote the
time instance and the scalar field value at the end of inflation.
Now in view of Eqs. (\ref{Friedman}) and (\ref{dotphi}), the
$e$-foldings number can be written as follows,
\begin{equation}\label{N}
    N=\frac{\kappa^2}{\lambda}\int_{\phi_f} ^{\phi_i} \frac{\mathcal{V}}{\mathcal{V}'} d \phi .
\end{equation}
Finally, let us recall at this point, that the current theory is a
rescaled scalar field theory with $\mathcal{R}^2$ quantum
corrections, thus the amplitude of the scalar perturbations of the
$f(\mathcal{R},\phi)$ theory at hand is not identical to the one
obtained by a simple scalar field theory. Thus it is necessary to
find an analytical expression for the amplitude of the scalar
perturbations for the theoretical framework at hand. Recall that,
the amplitude of scalar perturbations $\mathcal{P}_{\zeta}(k_*)$,
is defined as follows,
\begin{equation}\label{definitionofscalaramplitude}
\mathcal{P}_{\zeta}(k_*)=\frac{k_*^3}{2\pi^2}P_{\zeta}(k_*)\, ,
\end{equation}
and  one must evaluate it at CMB pivot scale $k_*=0.05$Mpc$^{-1}$.
The analytic expression for the amplitude of the scalar
perturbations for the $f(\mathcal{R},\phi)$ theory at hand is
\cite{Hwang:2005hb},
\begin{equation}\label{powerspectrumscalaramplitude}
\mathcal{P}_{\zeta}(k)=\left(\frac{k \left((-2
\epsilon_1-\epsilon_2-\epsilon_4) \left(0.57\, +\log \left(\left|
\frac{1}{1-\epsilon_1}\right| \right)-2+\log
(2)\right)-\epsilon_1+1\right)}{2 \pi  z}\right)^2\, ,
\end{equation}
with $z=\frac{(\dot{\phi} k) \sqrt{\frac{E(\phi
)}{f_{\mathcal{R}/\kappa^2 }}}}{\mathcal{H}^2 (\epsilon_3+1)}$,
and also for the $f(\mathcal{R},\phi)$ theory at hand, we have
$k=a\mathcal{H}$ at the first horizon crossing (for the CMB pivot
scale) and also at the first horizon crossing, where the amplitude
of the scalar perturbations must be evaluated, the conformal time
is $\eta=-\frac{1}{a\mathcal{H}}\frac{1}{-\epsilon_1+1}$. The
Planck data \cite{Planck:2018jri} constrain the amplitude of the
scalar perturbations to be
$\mathcal{P}_{\zeta}(k_*)=2.196^{+0.051}_{-0.06}\times 10^{-9}$.
Having developed the formalism for studying inflationary dynamics
in $f(\mathcal{R},\phi)$, we shall use the formalism to examine
the viability of several well known scalar field models which are
not compatible with the ACT data in the context of single scalar
field inflation, but in our present framework, can be compatible
with the ACT data.

\section{Models of Rescaled $\mathcal{R}^2$-corrected Gravity Compatible with the ACT data}

This section is devoted in the study of several well-known scalar
field inflationary theories, which are not compatible with the ACT
data, in the context of minimally coupled scalar field inflation.
We shall consider the rescaled $\mathcal{R}^2$ inflationary
versions of Higgs, hybrid, monomial and power-law inflationary
theories, and as we shall demonstrate, the resulting models are
compatible with the ACT data.

\subsection{Rescaled $\mathcal{R}^2$-corrected Hybrid Inflation}

\begin{figure}
\centering
\includegraphics[width=25pc]{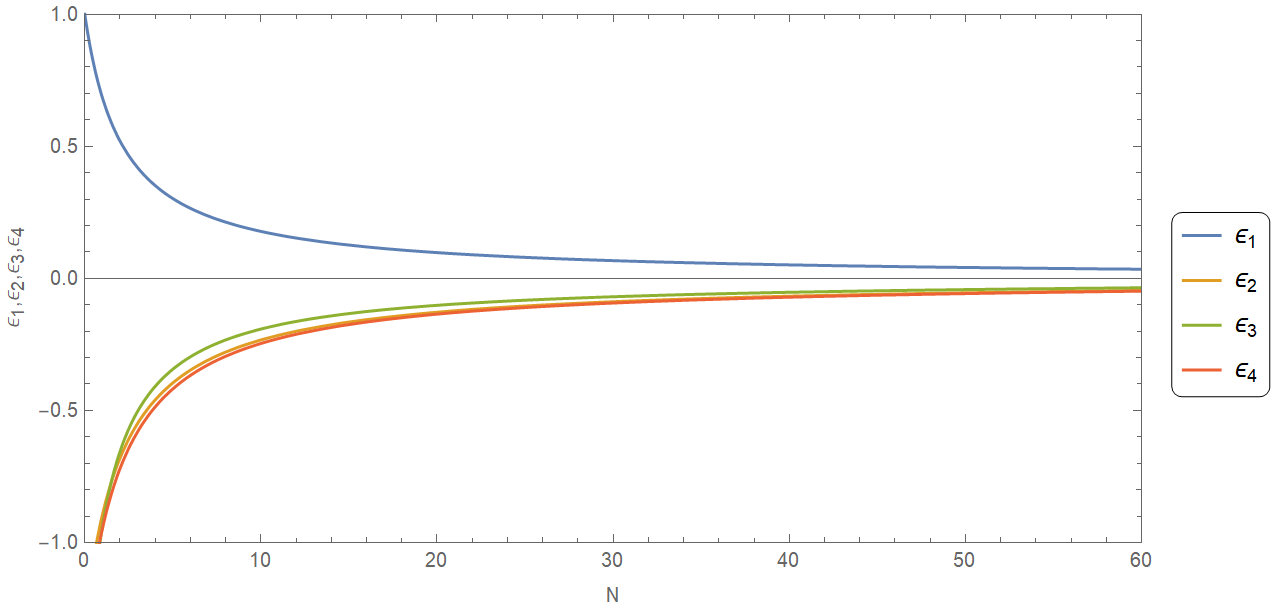}
\caption{The behavior of the slow-roll indices $\epsilon_1$,
$\epsilon_2$, $\epsilon_3$ and $\epsilon_4$, for the rescaled
hybrid inflation model in the $e$-foldings range $N=[0,60]$. We
used  the values of the free parameters $\mathcal{V}_0=8\times
10^{-12}$, $\beta=0.000004$, $\lambda=0.1$ and for $N$ in the
range $N=[0,60]$.}\label{refplot1}
\end{figure}

Let us apply the formalism of the previous section for a well
known inflationary potential, that of hybrid inflation
\cite{Linde:1993cn},
\begin{equation}\label{V1}
    \mathcal{V}(\phi)=\frac{\mathcal{V}_0}{\kappa^4}(\kappa \phi)^2,
\end{equation}
with $\mathcal{V}_0$ being a dimensionless free parameter. The
chaotic inflation model is not a viable model even in the single
scalar field inflation version, and it is incompatible with both
the Planck and ACT data. Substituting this scalar potential in
Eqs. (\ref{e1}), (\ref{e2}), (\ref{e3}),(\ref{ns}), (\ref{r}),
(\ref{N}), we obtain firstly the first slow-roll index
$\epsilon_1$,
\begin{equation}\label{e1V1}
    \epsilon_1=\frac{6 \kappa ^2 \lambda  M^2+4 \mathcal{V}_0}{3 \kappa ^4 M^2 \phi ^2}\, ,
\end{equation}
hence by solving the equation $\epsilon_1\simeq \mathcal{O}(1)$ we
obtain the value of the scalar field at the end of inflation,
$\phi_f=\frac{\sqrt{\frac{2}{3}} \sqrt{3 \kappa ^2 \lambda M^2+2
\mathcal{V}_0}}{\kappa ^2 M}$.
\begin{figure}
\centering
\includegraphics[width=25pc]{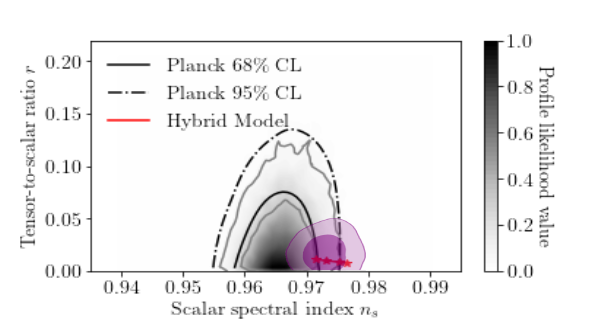}
\caption{The 2018 marginalized Planck likelihood curves, the ACT
constraints and the updated Planck constraints on the
tensor-to-scalar ratio, versus the rescaled
$\mathcal{R}^2$-corrected hybrid inflation model for
$\mathcal{V}_0=8\times 10^{-12}$, $\beta=0.000004$, $\lambda=0.1$,
and $N$ in the range $N=[50,60]$.}\label{hybrid}
\end{figure}
Accordingly, we obtain easily the second slow-roll index
$\epsilon_2=\frac{6 \kappa ^2 (\lambda -1) M^2+4 \mathcal{V}_0}{3
\kappa ^4 M^2 \phi ^2}$, and $\epsilon_3$ reads,
$\epsilon_3=\frac{4 \mathcal{V}_0 \left(3 \kappa ^2 \lambda  M^2+2
\mathcal{V}_0\right)}{-27 \kappa ^4 \lambda ^2 M^4+6 \kappa ^2 M^2
\mathcal{V}_0 \left(\lambda -\kappa ^2 \phi ^2\right)+4
\mathcal{V}_0^2}$. We proceed to finding the value of the scalar
field at the beginning of the inflationary era, because the
expression for the slow-roll index $\epsilon_4$ is too complicated
to quote here. Using Eq. (\ref{N}) and the value of the scalar
field at the end of inflation, we easily get,
$\phi_i=\frac{\sqrt{\frac{2}{3}} \sqrt{3 \kappa ^2 \lambda M^2+6
\kappa ^2 \lambda  M^2 N+2 \mathcal{V}_0}}{\kappa ^2 M}$. Now we
can easily confront the theory with the ACT data and a simple
analysis can reveal that this 3 parameter model can easily be
fitted within the ACT data. We take that $M=\beta/\kappa$ and for
$\mathcal{V}_0=8\times 10^{-12}$, $\beta=0.000004$ and
$\lambda=0.1$, for $N=55$ $e$-foldings we obtain
$n_{\mathcal{S}}=0.974329$, $r=0.00924384$ and also
$\mathcal{P}_{\zeta}(k)= 2.1535 \times 10^{-9}$. Hence the model
is compatible with the ACT data and the Planck data, including the
updated constraints on the tensor-to-scalar ratio. To have a
clearer picture of the compatibility of the model with the
observational data, in Fig. \ref{hybrid} we confront the model
with the 2018 marginalized Planck likelihood curves, the ACT
constraints and the updated Planck constraints on the
tensor-to-scalar ratio, for $\mathcal{V}_0=8\times 10^{-12}$,
$\beta=0.000004$ and $\lambda=0.1$, for $N$ chosen in the range
$N=[50,60]$. As it can be seen, the model can be well fitted
inside the observational data and also it is a strong gravity
model, meaning that it requires stronger gravity than the
Einstein-Hilbert gravity for it to be an ACT compatible model. One
can also check that the approximation of Eq. (\ref{taylorphi})
holds well true during the whole inflationary era, since at the
beginning of inflation we have (for $\phi=\phi_i$)
$\frac{2\kappa^2 \dot \phi^2}{M^2}\Big{|}_{\phi=\phi_i}=\frac{4
\lambda  \mathcal{V}_0}{3 \beta ^2}$ so for $\mathcal{V}_0=8\times
10^{-12}$, $\beta=0.000004$ and $\lambda=0.1$, $N=50$ we get
$\frac{2\kappa^2 \dot \phi^2}{M^2}\Big{|}_{\phi=\phi_i}=0.0666667$
while at the end of inflation when $\phi=\phi_f$ we have again
$\frac{2\kappa^2 \dot \phi^2}{M^2}\Big{|}_{\phi=\phi_i}=\frac{4
\lambda  \mathcal{V}_0}{3 \beta ^2}$ and for
$\mathcal{V}_0=8\times 10^{-12}$, $\beta=0.000004$ and
$\lambda=0.1$, we get $\frac{2\kappa^2 \dot
\phi^2}{M^2}\Big{|}_{\phi=\phi_f}=0.0666667$, so the value of the
quantity is small (and constant, but this is not important since
it is a model dependent feature) for the whole inflationary era.
It is also vital to examine the values of the slow-roll indices
for the whole duration of the inflationary era, in order to
determine whether the slow-roll condition breaks. In Fig.
\ref{refplot1} we present the behavior of the slow-roll indices
$\epsilon_1$, $\epsilon_2$, $\epsilon_3$ and $\epsilon_4$ for
$\mathcal{V}_0=8\times 10^{-12}$, $\beta=0.000004$, $\lambda=0.1$
and for $N$ in the range $N=[0,60]$. As it can be seen, the
slow-roll indices satisfy the slow-roll constraint $\epsilon_i\ll
1$, $i=1,2,3,4$ for this model, for the whole range of the
$e$-foldings number, except the last $e$-foldings of inflation
where their values increase towards unity, as expected. Note that
the last stages of inflation occur near $N=0$, as is known in
inflationary theories.

\subsection{Rescaled $\mathcal{R}^2$-corrected Monomial Inflation}

\begin{figure}
\centering
\includegraphics[width=25pc]{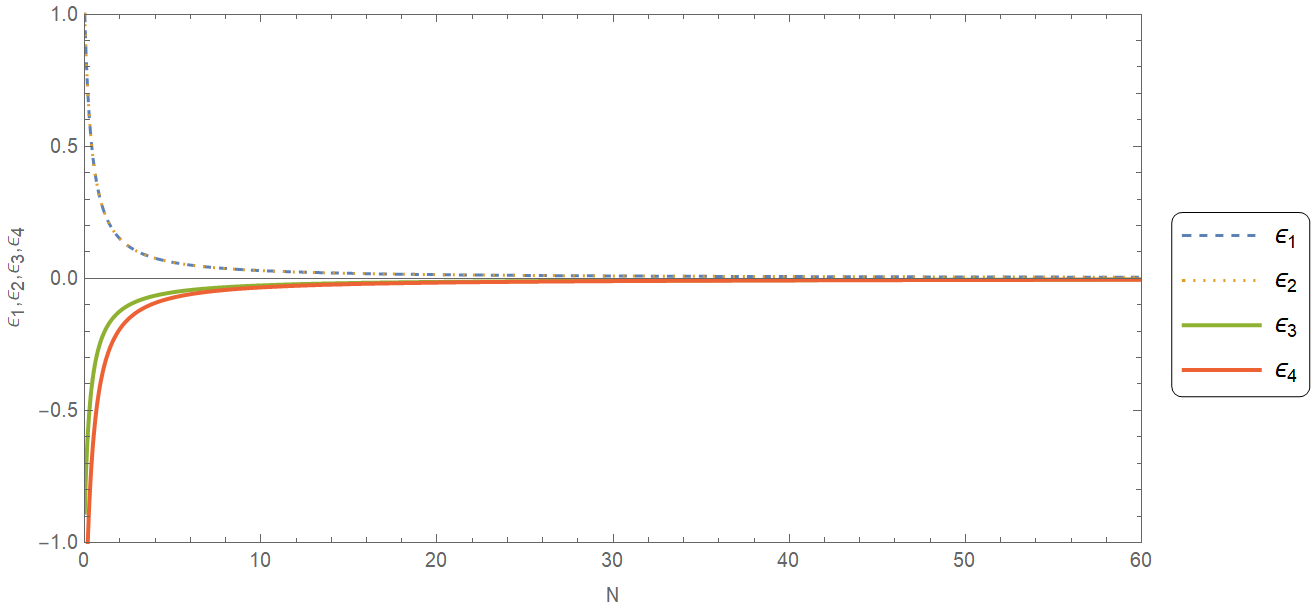}
\caption{The behavior of the slow-roll indices $\epsilon_1$,
$\epsilon_2$, $\epsilon_3$ and $\epsilon_4$, for the rescaled
monomial inflation model in the $e$-foldings range $N=[0,60]$. We
used  the values of the free parameters
$\mathcal{V}_0=0.000000001$, $\beta=0.000001$ and $\lambda=16.11$,
and $\delta=-0.45$ and for $N$ in the range
$N=[0,60]$.}\label{refplot2}
\end{figure}

Let us now consider another interesting inflationary scenario,
that of monomial inflation,
\begin{equation}\label{V12}
    \mathcal{V}(\phi)=\frac{\mathcal{V}_0 (1-\delta  (\kappa  \phi ))}{\kappa ^4},
\end{equation}
with $\mathcal{V}_0$ and $\delta$ being again dimensionless free
parameters. Substituting this scalar potential in Eqs. (\ref{e1}),
(\ref{e2}), (\ref{e3}),(\ref{ns}), (\ref{r}), (\ref{N}), we obtain
the first slow-roll index $\epsilon_1$,
\begin{equation}\label{e1V12}
    \epsilon_1=-\frac{\delta ^2 \left(\delta ^2 \mathcal{V}_0-6 \kappa ^2 \lambda  M^2 (\delta  \kappa  \phi -1)\right)}{12 \kappa ^2 M^2 (\delta  \kappa  \phi -1)^3}\, ,
\end{equation}
hence by solving the equation $\epsilon_1\simeq \mathcal{O}(1)$ we
obtain the value of the scalar field at the end of inflation,
which is too lengthy to quote it here.
\begin{figure}
\centering
\includegraphics[width=25pc]{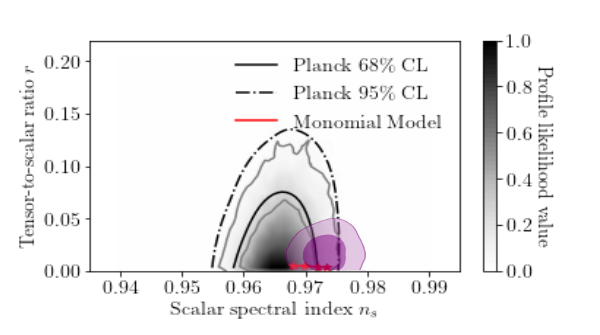}
\caption{The 2018 marginalized Planck likelihood curves, the ACT
constraints and the updated Planck constraints on the
tensor-to-scalar ratio, versus the rescaled
$\mathcal{R}^2$-corrected monomial inflation model for
$\mathcal{V}_0=0.000000001$, $\beta=0.000001$, $\lambda=16.11$,
and $\delta=-0.45$, and $N$ in the range
$N=[50,60]$.}\label{monomial}
\end{figure}
Accordingly, we easily obtain the second slow-roll index
$\epsilon_2=-\frac{\delta ^2 \left(\delta ^2 \mathcal{V}_0-6
\kappa ^2 \lambda  M^2 (\delta  \kappa  \phi -1)\right)}{12 \kappa
^2 M^2 (\delta  \kappa  \phi -1)^3}$, and the rest of the indices
and the value of the scalar field at the beginning of inflation.
Having these at hand, we can easily confront the theory with the
ACT data and a simple analysis can reveal that this 4 parameter
model can easily be fitted within the ACT data. For
$\mathcal{V}_0=0.000000001$, $\beta=0.000001$ and $\lambda=16.11$,
and $\delta=-0.45$ for $N=60$ $e$-foldings we obtain
$n_{\mathcal{S}}=0.973477$, $r=0.003905$ and also
$\mathcal{P}_{\zeta}(k)= 2.14609 \times 10^{-9}$. Hence this model
too is compatible with the ACT data and the Planck data, including
the updated constraints on the tensor-to-scalar ratio. In Fig.
\ref{monomial} we confront the model with the 2018 marginalized
Planck likelihood curves, the ACT constraints and the updated
Planck constraints on the tensor-to-scalar ratio, for
$\mathcal{V}_0=0.000000001$, $\beta=0.000001$ and $\lambda=16.11$,
and $\delta=-0.45$, for $N$ chosen in the range $N=[50,60]$. As it
can be seen, this model too can be fitted inside the observational
data, not as good as the hybrid inflation however. Also note that
this model requires a weaker gravity from the Einstein-Hilbert
gravity in order to be a viable model and its overall viability is
rated lower than the hybrid inflation one. Now the drawbacks of
this model. One can check in this case that the approximation of
Eq. (\ref{taylorphi}) does not hold true during the the
inflationary era. Let us see why, so at the beginning of inflation
for this model, with $\mathcal{V}_0=0.000000001$, $\beta=0.000001$
and $\lambda=16.11$, and $\delta=-0.45$, for $N=60$ we get
$\frac{2\kappa^2 \dot \phi^2}{M^2}\Big{|}_{\phi=\phi_i}=54.7185$
and at the end of inflation we have $\frac{2\kappa^2 \dot
\phi^2}{M^2}\Big{|}_{\phi=\phi_i}=584.055$, hence the
approximation of Eq. (\ref{taylorphi}) is violated, thus the
perturbative expansion does not hold true for this model at all.
It is intriguing that the only model which has inconsistencies
among the three models we will present, although ACT-compatible,
is the present model, which is a weak gravity one. Let us examine
the behavior of the slow-roll indices for this model too. In Fig.
\ref{refplot2} we present the behavior of the slow-roll indices
$\epsilon_1$, $\epsilon_2$, $\epsilon_3$ and $\epsilon_4$ for
$\mathcal{V}_0=0.000000001$, $\beta=0.000001$ and $\lambda=16.11$,
and $\delta=-0.45$ and for $N$ in the range $N=[0,60]$. As it can
be seen, the slow-roll indices satisfy the slow-roll constraint
$\epsilon_i\ll 1$, $i=1,2,3,4$ for this model too, for the whole
range of the $e$-foldings number, except the last $e$-foldings of
inflation where their values increase towards unity, as expected.
Note that the last stages of inflation occur near $N=0$, as is
known in inflationary theories.

\subsection{Rescaled $\mathcal{R}^2$-corrected Power-law Inflation}

\begin{figure}
\centering
\includegraphics[width=25pc]{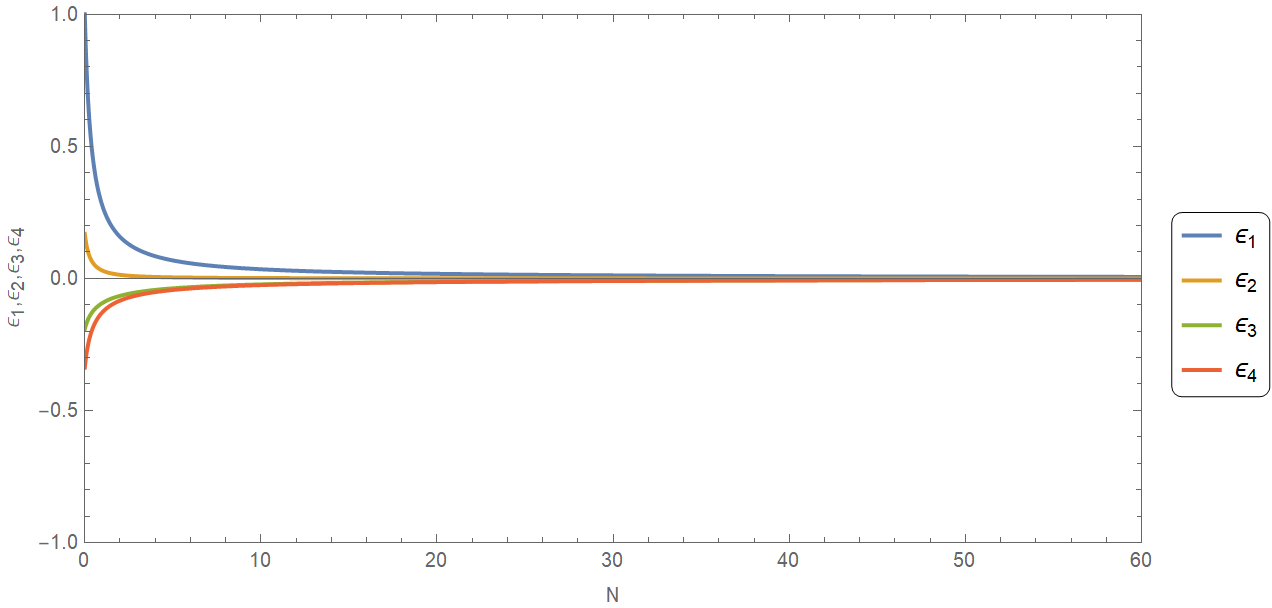}
\caption{The behavior of the slow-roll indices $\epsilon_1$,
$\epsilon_2$, $\epsilon_3$ and $\epsilon_4$, for the rescaled
power-law inflation model in the $e$-foldings range $N=[0,60]$. We
used  the values of the free parameters $\mathcal{V}_0=1.76\times
10^{-12}$, $\beta=0.0000023$ and $\lambda=0.48$ and for $N$ in the
range $N=[0,60]$.}\label{refplot3}
\end{figure}

Let us now consider a power-law inflation  \cite{Linde:1993cn},
\begin{equation}\label{V13}
    \mathcal{V}(\phi)=\frac{\mathcal{V}_0 (\kappa  \phi )^{4/3}}{\kappa ^4},
\end{equation}
where $\mathcal{V}_0$ being again a dimensionless free parameter.
Substituting this scalar potential in Eqs. (\ref{e1}), (\ref{e2}),
(\ref{e3}),(\ref{ns}), (\ref{r}), (\ref{N}), we obtain the first
slow-roll index $\epsilon_1$,
\begin{equation}\label{e1V3}
    \epsilon_1=\frac{8 \left(27 \kappa ^3 \lambda  M^2 \phi +8 \mathcal{V}_0 \sqrt[3]{\kappa  \phi }\right)}{243 \kappa ^5 M^2 \phi ^3}\, ,
\end{equation}
hence by solving the equation $\epsilon_1\simeq \mathcal{O}(1)$ we
obtain the value of the scalar field at the end of inflation,
which is too lengthy to quote it here, so we omit it.
\begin{figure}
\centering
\includegraphics[width=25pc]{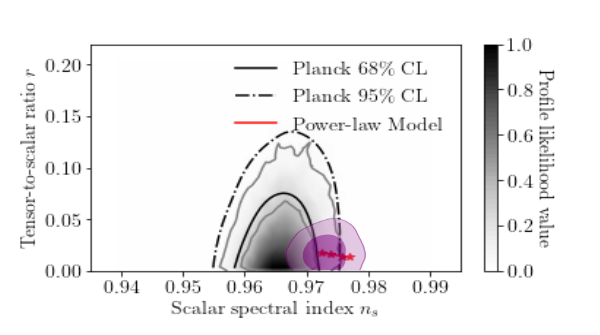}
\caption{The 2018 marginalized Planck likelihood curves, the ACT
constraints and the updated Planck constraints on the
tensor-to-scalar ratio, versus the rescaled
$\mathcal{R}^2$-corrected power-law inflation model for
$\mathcal{V}_0=8\times 10^{-12}$, $\beta=0.000004$, $\lambda=0.1$,
and $N$ in the range $N=[50,60]$.}\label{powerlaw}
\end{figure}
Accordingly, we obtain the second slow-roll index
$\epsilon_2=\frac{4 \left(27 \kappa ^2 (2 \lambda -1) M^2 (\kappa
\phi )^{2/3}+16 \mathcal{V}_0\right)}{243 \kappa ^2 M^2 (\kappa
\phi )^{8/3}}$, and accordingly we find the rest of the slow-roll
indices and the value of the scalar field at first horizon
crossing, but the analytic expressions are too lengthy to quote
here, so we omit them for brevity. So let us proceed to confront
the model with the ACT data and a some analysis can reveal that
this 3 parameter model can easily be fitted within the ACT data.
For $\mathcal{V}_0=1.76\times 10^{-12}$, $\beta=0.0000023$ and
$\lambda=0.48$, for $N=50$ $e$-foldings we obtain
$n_{\mathcal{S}}=0.972392$, $r=0.017771$ and also
$\mathcal{P}_{\zeta}(k)= 2.1596 \times 10^{-9}$. Therefore the
model is compatible with the ACT data and the Planck data,
including the updated constraints on the tensor-to-scalar ratio.
In Fig. \ref{powerlaw} we confront the model with the 2018
marginalized Planck likelihood curves, the ACT constraints and the
updated Planck constraints on the tensor-to-scalar ratio, for
$\mathcal{V}_0=1.76\times 10^{-12}$, $\beta=0.0000023$ and
$\lambda=0.48$, for $N$ chosen in the range $N=[50,60]$. As it can
be seen, the power-law model can be well fitted inside the
observational data. We need to mention that in this case too, the
model requires stronger gravity for it to be viable, compared with
the Einstein-Hilbert gravity. Let us also check the approximation
of Eq. (\ref{taylorphi}) if it holds true during the inflationary
era. At the beginning of inflation for $\mathcal{V}_0=1.76\times
10^{-12}$, $\beta=0.0000023$ and $\lambda=0.48$, for $N=50$ we get
$\frac{2\kappa^2 \dot \phi^2}{M^2}\Big{|}_{\phi=\phi_i}=0.0235934$
while at the end of inflation when $\phi=\phi_f$ for
$\mathcal{V}_0=1.76\times 10^{-12}$, $\beta=0.0000023$ and
$\lambda=0.48$, $N=50$, we get $\frac{2\kappa^2 \dot
\phi^2}{M^2}\Big{|}_{\phi=\phi_f}=0.11660$, hence the value of the
quantity $\frac{2\kappa^2 \dot \phi^2}{M^2}$ is small during the
whole inflationary era. Let us examine the behavior of the
slow-roll indices for this model too. In Fig. \ref{refplot3} we
present the behavior of the slow-roll indices $\epsilon_1$,
$\epsilon_2$, $\epsilon_3$ and $\epsilon_4$ for
$\mathcal{V}_0=1.76\times 10^{-12}$, $\beta=0.0000023$ and
$\lambda=0.48$ and for $N$ in the range $N=[0,60]$. As it can be
seen, the slow-roll indices satisfy the slow-roll constraint
$\epsilon_i\ll 1$, $i=1,2,3,4$ for this model too, for the whole
range of the $e$-foldings number, except the last $e$-foldings of
inflation, where the value of $\epsilon_1$ solely increases
towards unity, as expected. Note in this case too, that the last
stages of inflation occur near $N=0$, as is known in inflationary
theories.

\section{Conclusions}

The failure of the terrestrial experiments to verify any UV
completions of the Standard Model, makes the experimental
verification of them hard, if not impossible. However, imprints of
the Planck quantum era can be found on the inflationary
Lagrangian, and thus can be verified by the observational data.
The recent ACT data have excluded many single scalar field theory
models of inflation. In this work we presented a theoretical
framework that revives many of the ACT-incompatible models of
single scalar field inflation. The approach we adopted is simple,
and we assumed that the inflationary Lagrangian is controlled by
the synergy of single scalar field and curvature quantum
corrections. Specifically we introduced the theoretical framework
of rescaled $\mathcal{R}^2$-corrected minimally coupled scalar
field theory and we showed that it can easily be compatible with
the ACT data. We formulated the theory in such a way so that it
leads to a single scalar field theory in the Jordan frame and not
in a two scalar field theory in the Einstein frame. The attribute
of our approach is that the Jordan frame is the natural choice,
and the correct choice for physics, because in the Jordan frame,
baryon geodesics are free fall geodesics. We considered the first
two quantum corrections of the single scalar field Lagrangian,
linear $\sim \mathcal{R}$ and a quadratic term $\sim
\mathcal{R}^2$. The linear term leads to a rescaled
Einstein-Hilbert term $\sim \lambda \frac{\mathcal{R}}{16\pi G}$,
which in effect leads to a rescaled Newton's gravitational
constant $\frac{G}{\lambda}$. Thus gravity can theoretically be
stronger or weaker than present day's Newton's constant.
Intuitively, one expects gravity to be stronger primordially, but
this is just intuition imposed by the fact that spacetime was tiny
primordially, thus gravity should be much stronger. In most
inflationary contexts, gravity has the same strength today and
primordially, in our framework gravity is primordially stronger in
an effective field theory way, while as the spacetime expands, it
becomes Einstein-Hilbert gravity again. We examined three
mainstream minimally coupled single scalar field models of
inflation, which are incompatible with the ACT data, the hybrid
inflation, the monomial inflation and the power-law inflation. The
hybrid inflation in the context of rescaled
$\mathcal{R}^2$-corrected minimally coupled scalar field theory is
a 3-parameter model which is compatible with the ACT data and also
self-consistent, in its strong gravity limit. The power-law
inflation is again a 3-parameter model, consistent and also
compatible with the ACT data in its strong gravity regime. The
monomial inflation is a 4-parameter model, compatible with the ACT
data in its weak gravity regime, however it suffers from
theoretical inconsistencies. Interestingly enough, the only models
which generate an ACT-compatible inflationary era and also are
self-consistent, are the two strong gravity models.

An immediate extension of this work is to also consider
Einstein-Gauss-Bonnet terms. Such a framework was studied in
\cite{Odintsov:2020ilr} so work is in progress to also consider
the rescaled version of this theory in view of the ACT data.

\section*{Acknowledgments}

This research has been funded by the Committee of Science of the
Ministry of Education and Science of the Republic of Kazakhstan
(Grant No. AP26194585) (V.K. Oikonomou).

\end{document}